\documentstyle[twocolumn,aps,prl,epsf]{revtex}

\begin{document}
\draft
\title{Anholonomy and Geometrical Localization in Dynamical Systems }
\author{ Radha Balakrishnan$^a$ and Indubala I Satija$^b$}
\address{
(a)The Institute of Mathematical Sciences,Chennai 600 113, India
(b)Department of Physics, George Mason University, Fairfax, VA 22030}
\maketitle
\begin{abstract}
We characterize the geometrical and topological
aspects of a dynamical system by associating
a geometric phase with a
phase space trajectory.
Using the example of a nonlinear driven damped oscillator,
we show that this phase is resilient to fluctuations, responds to all bifurcations
in the system, and also finds new geometric transitions.
Enriching the phase space description
is a novel phenomenon of ``geometrical localization''
which manifests itself as a significant deviation
from planar dynamics over a short time interval.
\end{abstract}

\narrowtext

Anholonomy is a  phenomenon
that arises if a  quantity  fails to recover  its original value,
when  the parameters on which it depends are varied round a closed path.
First introduced by Berry\cite{berry} in quantum mechanics as
a geometric phase associated with
the wave function, it is now recognized that this phase can
emerge in classical systems as well.
The universality underlying this  concept has proven to be very useful
in a variety of systems\cite{wilczek,radha}.

In this Letter we present, for the first time, a  formalism which
uses this  concept to  quantify the geometric characteristics of
complex phase space trajectories of systems
exhibiting chaotic dynamics. Here, the root of the anholonomy
lies in the {\it non-planarity} of the trajectories.
Using the example of a nonlinear driven damped oscillator,
we demonstrate that the geometric phase is a robust order parameter
for the system, which not only responds to all bifurcations but also {\it finds}
new  geometrical transitions. These include
geometrical resonances leading to both discontinuous as well as continuous
changes in the geometry of periodic and strange attractors.
Our formalism integrates phase space dynamics and its
geometrical aspects by including information about
the non-planarity of the trajectory.
Underlying this enrichment
is the novel phenomenon of {\it geometrical localization}:
 the temporal localization of the geometrical variables,
when the deviation from planar dynamics occurs over small time intervals.

Although our formalism can be generalized to higher dimensions,
we consider (as an illustrative example)
a dynamical system described by three
first-order coupled differential equations,
\begin{equation}
{\bf r}_{t}= {\bf v}({\bf r}, \alpha)\,;\,\,\,\,
{\bf r}(t) =(x(t)\,,\, y(t)\,,\, z(t))\,\,,
\label{dynsys}
\end{equation}
where the subscript $t$ denotes the time-derivative,
 $(x, y, z)$  are the variables of the 3D phase space,
${\bf v}$ is a  given function of these variables,
and $\alpha$ is a control parameter.
Driven damped nonlinear oscillators, Duffing's equations for
nonlinear mechanical vibrations,
the Lorenz equations describing 2D convection in a heated fluid,
nonlinear electronic circuit equations, etc.,
are some well-known examples\cite{drazin} of Eq. (\ref{dynsys}).

A  phase trajectory of the dynamical system of Eq. (\ref{dynsys})
can be viewed as a
space curve generated by the three-dimensional
vector ${\bf r}(t)$
parametrized by the time $t$. From Eq. (\ref{dynsys}),
we have $|{\bf r}_t|=|{\bf v}|=v =s_{t}$, giving $s(t) = \int v~dt$
 as the arc length on the space curve.
Thus the  unit tangent vector is given by
 ${\bf T}={\bf r}_{t}/v$. In accordance with the usual formalism \cite{struik}
 for a space curve, we define  the orthogonal right-handed
triad of unit vectors $({\bf T},{\bf N},{\bf B})$,
where ${\bf N}$ and ${\bf B}$ denote, respectively,
the normal and binormal unit vectors on the curve. The
Frenet-Serret (FS) equations can be written in terms of the variable $t$
as
\begin{equation}
{\bf T}_{t} = v\kappa {\bf N}\,,\,
{\bf N}_{t} = - v\,\kappa \,{\bf T} + v\,\tau \,{\bf B}\,,\,
{\bf B}_{t} = - v\,\tau \,{\bf N}\,,
\label{fseqns}
\end{equation}
where the curvature $\kappa$ and the torsion $\tau$
are functions of $s$ and determine the local geometry of the trajectory.
It can be shown that\cite{struik}
\begin{eqnarray}
\kappa(s)&=&|{\bf r}_{t} \times {\bf r}_{tt}|/|{\bf r}_{t}|^{3}\,,\\
\tau(s) &=& {\bf r}_{t} \cdot({\bf r}_{tt}
\times {\bf r}_{ttt}) /|{\bf r}_{t}\times {\bf r}_{tt}|^2\,.
\label{geomvar}
\end{eqnarray}
Intuitively, the curvature measures the
departure of a curve from a straight line, while
the torsion measures its non-planarity.

To understand how the anholonomy of a trajectory
arises\cite{dand}, we rewrite Eqs. (\ref{fseqns}) as
${\bf F}_t= \xi \times {\bf F}$,
where ${\bf F}$ stands for
${\bf T},\,{\bf N},$ or ${\bf B},$ and
$\xi =-v\,\kappa {\bf B} + v\tau {\bf T}$.
This shows that as one moves on
the trajectory ${\bf r}(t)$, the
FS triad $({\bf T},\,{\bf N},\,{\bf B})$
rotates with angular velocities $v\,\kappa$ and $v\,\tau$
around ${\bf B}$ and
${\bf T}$ respectively. In the $({\bf N},{\bf B})$ plane, we may
introduce two orthogonal unit vectors
 ${\bf u}$ and ${\bf w}$ such that the triad
$({\bf T},\, {\bf u}, \,{\bf w})$ does not rotate around ${\bf T}$.
This is achieved by using a Fermi-Walker parallel transport
of any vector ${\bf P}$ moved along the curve
according to $\delta {\bf P}/\delta {t}=
{\bf P}_t-v\,\kappa ({\bf B}\times{\bf P})$. Thus,
as one moves from $t=0$ to $t=T$, a geometric phase
$\Phi_{T}=\int_{0}^{T} v\,\tau\, dt
= \int_{\gamma (s)} \tau\,ds$
develops between its natural frame $({\bf N},\,{\bf B})$ and
the ``non-rotating frame'' $({\bf u},\,{\bf w})$.
The anholonomy $\Phi_{T}$ can also be interpreted in another way:
Representing the rotation of the triad using
Euler angles $(\theta,\,\varphi, \,\psi)$
yields\cite{dand} $\Phi_{T} = 2\pi-\int \sin\, \theta \,d\theta \int d\varphi$.
For a periodic trajectory, the second term is just the solid angle subtended
by the area enclosed by the closed path $\gamma(s)$
traced out by the tangent indicatrix \cite{struik} on the unit sphere $S^{2}$.
The same result holds
good for a non-periodic trajectory as well,
 since it can  always be closed using a geodesic\cite{sam} on the sphere.

As is clear, the anholonomy $\Phi_{T}$ of a trajectory also characterizes its non-planarity.
Since both periodic as well as chaotic
trajectories occur in general, a more useful quantity is the
geometric phase per unit time, defined over a long time $T$,
\begin{equation}
\phi_{T} \equiv \frac{\Phi_{T}}{T} =
\frac{1}{T}
\int_{0}^{T} v\,\tau\, dt
\label{gpaverage}
\end{equation}
where $T$ is much larger than all the natural time scales in the
problem. $\phi_{T}$ as defined above
can also be thought of as the
mean angular
velocity of the FS triad.
As we show below, it turns out to be
a good order parameter for both periodic and
chaotic trajectories.

We now apply this formulation to
a driven, damped, bilinear oscillator\cite{bs}
characterized by two {\it different} frequencies
$\omega_{1,2}$ for positive
and negative displacements respectively:
\begin{equation}
x_{tt}+2\beta \,x_{t}= -\omega_{1,2}^{2}\,x + f \,\cos \,\omega t\,,\,
(\beta > 0)\,.
\label{blosc}
\end{equation}
This can be written in the form of Eq. (1) by choosing ${\bf r}=(x,y,z)=(x, x_t, x_{tt})$.
For $\beta =0,\,f = 0\,$, the oscillator has a frequency $\omega_{\rm
bl}= 2\omega_{1}\omega_{2}/(\omega_{1}+\omega_{2})$.
The system is piecewise
linear, and analytic solutions can be obtained for $x > 0$ and $x <0$.
The discontinuity at the origin makes it essentially nonlinear,
$\alpha=\omega_2/\omega_1$ being the
nonlinearity parameter.
Without loss of generality, we
choose the units of $t$ and $x$ such that $\omega=1$ and $f=1$.
The two key control parameters are then $\omega_1$ and $\omega_2$,
or, equivalently, $\alpha$ and $\omega_1$.
The results to be presented are for fixed $\omega_1=1.5$,
although we have investigated the full
two-dimensional parameter space\cite{www}.
For a linear damped
driven oscillator ($\omega_1=\omega_2$),
it is easy to show that $\tau$ calculated using
Eq. (\ref{geomvar}) is identically zero and hence the
attractor is  a planar limit cycle.

Our numerical routine uses the analytical
solution for $x(t)$ in each half of the
phase space. Starting with an initial state $x >0\,,\,x_t = 0$
(so that the oscillator frequency is $\omega_1$),
we use very small time increments to determine precisely
the time and the velocity when $x$ reaches $0$.
These are used as the new initial conditions for dynamics
in the region $x < 0$ with oscillator frequency $\omega_2$, and the
process is repeated. The availability of an
analytic solution is particularly useful
for computing the torsion of the phase trajectory using
Eq. (\ref{geomvar}).

We have carried out a detailed exploration \cite{www} of the system in
the parameter space, and
calculated  the local phase space variable
$x$,  the local geometric variable $\tau$,
the long-time average  anholonomy $\phi_{T}$, as well as
$\phi_1\,,$ the average anholonomy over a {\it single} period of the driving
force.
\begin{figure}
\begin{center}
\leavevmode
\epsfxsize=3in
\epsfbox{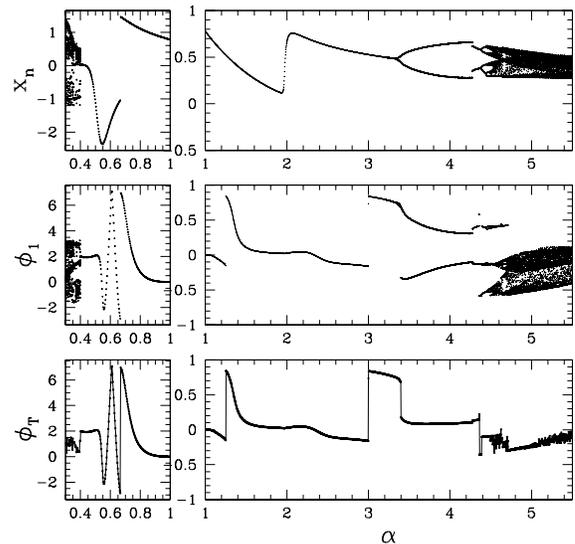}
\end{center}
\protect\caption{ Bifurcations and geometrical resonances:
The attractor at Poincar\'e points $x_{n}$ and the geometric
phase factors $\phi_{1}$ and ${\phi_{T}}$
as the nonlinearity parameter $\alpha$ is varied.
Plots for $\alpha > 1$ and $\alpha < 1$ are shown
separately to highlight the differences in the magnitude of the anholonomy.}
\label{fig1}
\end{figure}

 The bilinear oscillator exhibits period-doubling transitions, followed by
chaotic
dynamics which is reflected  by  $\phi_1$ in addition to phase space variables.
A close look at Fig. ~1 shows that all changes
in the attractor in phase space are reflected in $\phi_T$
as well, either as a
discontinuity in $\phi_T$ or as a change in its slope
$\partial \phi_T/\partial \alpha$.
For example, at $\alpha = 2/3$ ($\omega_2=1$) and $\alpha \approx 4.25$
there is a sudden change in the Poincar\'e plot $x_n$ (due to coexisting
attractors changing their basin
of attraction) which is
accompanied by a jump in $\phi_T$. A sequence of period-doublings
at $\alpha \approx 3.4, 4.25, 4.7...$ as well as a resonance at $\alpha \approx
2.0$, where $\omega_{bl}=\omega$,
are accompanied by changes in
$\partial \phi_T/\partial \alpha$.

A striking feature is the emergence of a series of
new {\it geometric} transitions of the attractor that are not captured by
the standard bifurcation diagram for $x_n$.
While $x_n$ varies smoothly, $\phi_1$ and $\phi_T$ show a jump at $\alpha \approx 1.25,
3.0, 3.4\,$,
reminiscent of a first-order phase transition.
In contrast, at $\alpha \approx 0.61\,,$
$x_{n}$ and $\phi_T$  are both  continuous, but
$\partial \phi_T/\partial \alpha$ has a jump, reminiscent of a
second-order phase transition. These transitions are robust, as the threshold
parameter values
for both first and second order transitions
lie on a smooth one-dimensional curve in two-dimensional parameter
space\cite{www}.
Based on our detailed numerical studies, there is ample justification
for regarding $\phi_T$ as an
order parameter for such geometric transitions.

The first order transitions
result in a discontinuous change in $\phi_1$ by $\omega$ ( which is set to unity here) in only
one of the branches of $x_n$.
As a result, we see discontinuities in $\phi_{T}$ by  exactly $1/n$,
where $n$ is the period of the attractor in units of $2\pi/\omega$.
Therefore, the discontinuities in $\phi_T$
correspond to a full rotation of the FS triad
by $2\pi$ in a full period of the driving force.
These transitions can thus be viewed as a resonance
effect between
the average rotation frequency of the FS triad
and the driving frequency.
Higher order resonances where the FS triad rotates by $2\pi m$ ($m =$ any integer)
were also frequently observed\cite{www} for $\alpha < 1$.
In the neighborhood of a second-order transition, the mean anholonomy is very large
as the FS triad executes many rotations (not necessarily an integer
number) during one period of the driving.
Such sudden responses in the geometry of the trajectories that are not captured by usual
bifurcation diagrams will be referred to as {\it geometrical resonances} (GR).

\begin{figure}
\begin{center}
\leavevmode
\epsfxsize=3in
\epsfbox{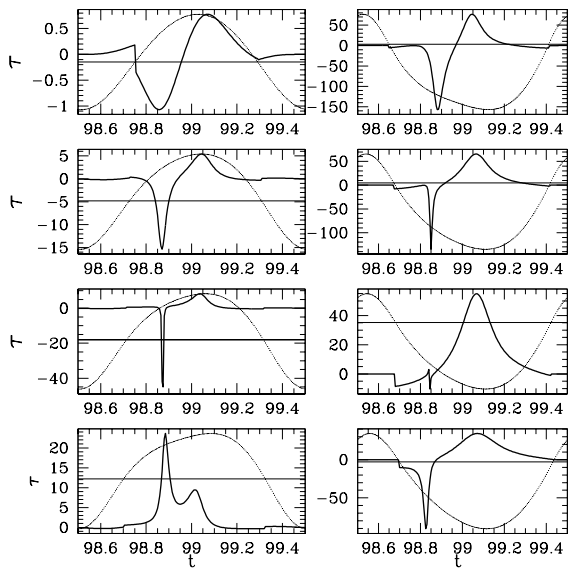}
\end{center}
\protect\caption{Geometrical localization near first-order
(left) and second-order(right)
geometrical transitions.
Time series for $\tau$ (thick lines) and
$x$ (dashed lines; full $x$-scale not shown), shown over one driving cycle.
The horizontal line corresponds to $x = 0$.
Unlike $x$ and $x_{t}$, $\tau$ is sensitive to the discontinuity
in  the acceleration  at $x=0$.
>From top to bottom: on the left,
$\alpha=1.1, \,1.2,\, 1.25, \, 1.3\,$;
on the right, $\alpha=0.66,\, 0.62,\, 0.61,\, 0.59$.}
\label{fig2}
\end{figure}

Further insight into these intriguing phenomena
can be gained by studying the local dynamics
of both the phase space and geometrical
variables as shown in Fig. 2 .
As we discuss below, the key to the various effects
is the localization characteristics
underlying the torsion $\tau$, which we will refer to as the $\tau$-mode.
We first discuss the dynamics near the first-order transition,  given on the left
in  Fig. 2.
An interesting aspect is the {\it temporal localization} of the $\tau$-mode
corresponding to
large geometrical changes in very short time intervals, as the nonlinearity
 increases from $\alpha=1.1$. At a critical parameter value, it becomes a
$\delta$-function and then {\it flips}, i.e., the localized $\tau$ changes its sign,
after which the  $\tau$-dynamics  that was ``bi-directional'' (i.e., $\tau$ had both signs)
becomes  ``unidirectional'', with the
FS triad rotating in one direction alone, as  can be inferred from the figure.
The geometrical transition appears to be linked to subtle changes in the
dynamics. Here the local curvature vanishes as
${\bf r}_t$ and ${\bf r}_{tt}$ become
collinear. In terms of phase space variables,
the force and its two derivatives are very small
and the particle is
subjected to a small constant force
for a long time interval.

One of the interesting aspects of a first-order
GR transition is the fact that
it is often associated with a period-doubling
bifurcation; in that case
it is a ``twin" transition,
where a jump in $\phi_1$ before period-doubling
is always accompanied  by another jump (of identical
magnitude but opposite sign) in $\phi_1$
after the period-doubling. The twin aspect of the GR was seen throughout the two-dimensional
parameter space, and may be universal.

In contrast to the  first-order case, the dynamics near a second-order transition
corresponds to an enhancement in the localization length of the $\tau$-mode.
as described on the right in Fig. 2.
In addition to this delocalization, the mode also becomes predominantly unidirectional (positive $\tau$)
due to the disappearance of the strongly localized (negative) $\tau$. After this stage, the mode becomes
 bi-directional again.
In view of the absence of "flipping", $\phi_T$ remains continuous,
while its derivative exhibits a discontinuity.

\begin{figure}
\begin{center}
\leavevmode
\epsfxsize=3in
\epsfbox{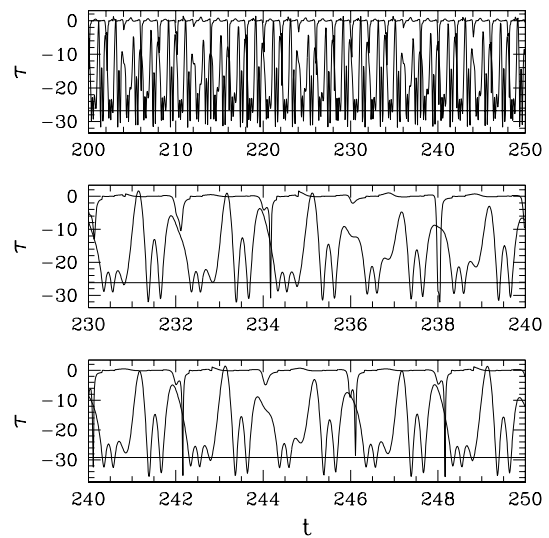}
\end{center}
\protect\caption{Time series
for $\tau$ (thick line) and $x$ (thin line) in the chaotic state for
$\alpha=5$. 
The middle and the bottom plots show the blowups. }
\label{fig3}
\end{figure}

Localization of $\tau$ implies that the
non-planarity of the attractor is
significant only in short intervals of time.
This nonlinearity-induced
localization of the geometrical
variables when the phase space variables
continue to exhibit oscillatory dynamics is
an interesting new phenomenon that we will refer to
as {\it geometrical localization}. In addition to their existence near GRs, localized
$\tau$-modes are seen frequently as sudden,
seemingly  random bursts in the
chaotic time series as clearly seen in Fig. 3. They are  tied to very subtle changes in the dynamics associated with
the derivatives of $x$.

We had stated  earlier that $\phi_{T}$ is an appropriate
order parameter even
for dynamics on a chaotic attractor. This is borne out by the
fact that $\phi_{T}$ converges to a well-defined asymptotic
value for very large sampling times $t$, as shown in Fig. 4.

\begin{figure}
\begin{center}
\leavevmode
\epsfxsize=3in
\epsfbox{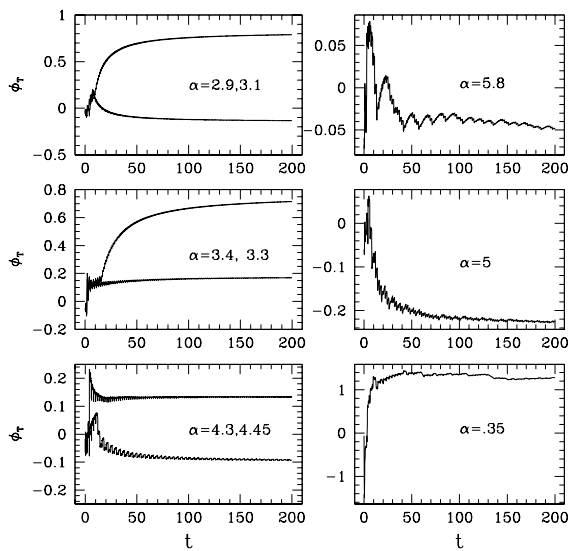}
\end{center}
\protect\caption{ Convergence of $\phi_{T}$ for the attractor ( i.e. large $t$)
 for various values of $\alpha$
in the periodic (left) and chaotic (right) dynamics.
The two curves in each of the plots on the left show
$\phi_T$ before and after GRs, respectively.}
\label{fig4}
\end{figure}
The average anholonomy fluctuates before converging to a fixed value,
reminiscent of the critical fluctuations
of an order parameter near a thermodynamic phase transition.
The fact that $\phi_T$ converges to a unique asymptotic limit even for chaotic
trajectories establishes the resilient aspect of the geometric phase
in the classical bilinear oscillator.
This {\it tolerance to fluctuations} may be a universal characteristic of
geometric phases in both quantum and classical systems which has led
to the exciting possibility of geometric quantum computation\cite{nature}.

It is interesting to note that, by defining a complex vector
  ${\bf M}= ({\bf N}+i{\bf B})/\sqrt {2}$,
 a short calculation using the FS equations (\ref{fseqns}) shows that
 the {\it classical} geometric phase
  $\Phi_{T}= \int_{\Gamma(s)} \tau ds = i\int_{\Gamma(s)}
{\bf M}^{*}\cdot {\bf M}_{s} ds $.
 This expression  has exactly the same form as the {\it quantum}
geometric phase found by  Berry \cite{berry}, when  the classical
 vector ${\bf M}$ is  replaced by a quantum state  $|n({\bf R})\rangle$.
In view of the general robustness of the geometric phase,
we believe that our formalism that uses anholonomy may provide
 an alternative formulation for studying the classical-quantum correspondence.

In summary,  our space curve formulation of dynamical systems using
the FS equations
provides a new method which adds geometrical features
to the usual phase space description of the complex dynamics
of classically chaotic systems.
Non-planarity of the attractor is related to anholonomy,
which, in turn, is shown to  be an appropriate robust  order
parameter that characterizes the geometry of the attractors.
Adding to the richness of the bifurcation diagram
is the prediction of first and second-order transitions in the geometry
 of the attractor. These transitions are related to the localization characteristics
of the torsion. This suggests
 the novel concept of geometrical localization near a transition, which
emerges due to the fact that significant deviation
from planar dynamics
exists only over small time intervals. We thus relate
nonlinearity  not only  to anholonomy but
also to localization.

We believe that several of the features we have seen in our geometric
 description of the bilinear oscillator may
be generic and found  in other nonlinear systems as well.
These were seen throughout the
two-dimensional parameter space\cite{www}, and also
for other choices of phase space variables such as
${\bf r}=(x, \,x_t,\, t)$ and
$(x,\, x_t,\, \cos\,\omega t)$.
Our formulation can also be extended to $D\ge 4$.
For instance, when $D=4$, the  anholonomy can be found  from
 the  expression for the angular velocity of a
 tetrad that can be  appropriately defined for the system.
We hope that our study will stimulate a new line of
research relating geometry and dynamics
in nonintegrable systems in general.

The research of IIS is supported by National Science
Foundation Grant No. DMR~0072813.

\end{document}